\global\def\draftcontrol{0}

   \def\versionno{ dream }

\catcode`\@=11

\expandafter\ifx\csname draftcontrol\endcsname\relax\global\def\draftcontrol{0} 
\fi 

{\count255=\time\divide\count255 by 60 
\xdef\hourmin{\number\count255} 
\multiply\count255 by-60\advance\count255 by\time 
\xdef\hourmin{\hourmin:\ifnum\count255<10 0\fi\the\count255}} 
\def\draftdate{\number\month/\number\day/\number\year\ \ \ \hourmin } 

\newcommand\makepapertitle{\par

  \begingroup 
    \renewcommand\thefootnote{\@fnsymbol\c@footnote}%
    \def\@makefnmark{\rlap{\@textsuperscript{\normalfont\@thefnmark}}}%
    \long\def\@makefntext##1{\parindent 1em\noindent 
            \hb@xt@1.8em{%
                \hss\@textsuperscript{\normalfont\@thefnmark}}##1}%
     \newpage 
     \global\@topnum\z@   
     \@makepapertitle 
     \thispagestyle{empty}\@thanks 
  \endgroup 
  \setcounter{footnote}{0}%
  \global\let\thanks\relax 
  \global\let\makepapertitle\relax 
  \global\let\@makepapertitle\relax 
  \global\let\@thanks\@empty 
  \global\let\@author\@empty 
  \global\let\@date\@empty 
  \global\let\@title\@empty 
  \global\let\title\relax 
  \global\let\author\relax 
  \global\let\date\relax 
  \global\let\and\relax 
  \def\version{\let\version\@version\@gobble} 
} 
\def\@makepapertitle{%
  \newpage 
   \ifnum\draftcontrol=1 {} 
   \version\versionno 
   \vskip 5em%
   \else 
   \hfill\hbox to 3cm {\parbox{4cm}{\@pubnum}\hss}%
   \vskip 5em%
   \fi 
   \begin{center}%
   \let \footnote \thanks 
      {\hskip -0\textwidth \hbox to 1\textwidth%
        {\centerline{\Large\bf{\noindent\@title}}}}%
     \vskip 2em%
     {\normalsize
       \lineskip .5em%
       \begin{tabular}[t]{c}%
         \@author 
       \end{tabular}\par}%
     \vskip 1.5em%
     {\@bstract}%
     \end{center}%
     \vfill
     \@date%
     \vskip 1.5em%
   \par 
} 

\gdef\@pubnum{} 
\def\pubnum#1{%
  \gdef\@pubnum{#1}} 

\gdef\@bstract{} 
\def\Abstract#1{%
  \gdef\@bstract{%
   \parbox{\textwidth-0pc}{%
   \centerline{\bf Abstract}\penalty1000 
   \noindent
   \renewcommand\baselinestretch{1.0} 
   {#1}}} 
} 

\gdef\@email{}
\def\email#1{%
   \gdef\@email{%
   Email: {\tt #1}}
}

\def\ps@paper{\let\@mkboth\@gobbletwo%
     \ifnum\draftcontrol=1 
        \def\@oddfoot{\hbox to \textwidth{\tiny \versionno \hfil\tiny\draftdate}%
        \hskip -\textwidth \hbox to \textwidth{\hfil\rm\thepage\hfil}}%
     \else\def\@oddfoot{\hbox to \textwidth{\hfil\rm\thepage\hfil}} 
     \fi 
     \let\@evenfoot\@oddfoot 
} 

\def\body{\clearpage 
          \pagestyle{paper} 
        } 
\newenvironment{acknowledgments}{%
\vskip 3.25ex 
\noindent {\bf Acknowledgments} 
} 

\def\@version#1{\ifnum\draftcontrol=1 
\typeout{}\typeout{#1}\typeout{} 
\vskip3mm\centerline{\hbox{\fbox{\normalsize{\tt DRAFT -- #1 -- } 
                   {\draftdate}}}}\vskip3mm 
\fi} 
\let\version\@version 
\long\def\eqlabel#1{\ifnum\draftcontrol=1 
                    \tag@false  
                    \tag*{(\theequation) \hbox to -0.2cm{\hspace{0cm}\small{#1}\hss}} 
                    \refstepcounter{equation}  
                    \edef\@currentlabel{\theequation} 
                    \ltx@label{#1}          
                    \else 
                    \label{#1} 
                    \fi 
                    } 
\let\st@bibitem\@bibitem 
\let\st@lbibitem\@lbibitem 
\ifnum\draftcontrol=1 
  \def\@bibitem#1{%
    \st@bibitem{#1}\a@@label{#1}\ignorespaces} 
  \def\@lbibitem[#1]#2{%
    \st@lbibitem[#1]{#2}\a@@label{#2}\ignorespaces} 
  \def\a@@label#1{%
    \gdef\a@lab{\smash{\normalfont\small#1}} 
    \ifvmode 
      \if@inlabel 
        \global\setbox\@labels\hbox{%
          \llap{\a@lab\let\a@lab\relax 
                \kern\@totalleftmargin\kern\marginparsep}%
          \box\@labels}%
      \fi 
    \fi} 
\fi 

\documentclass[12pt,letterpaper]{article} 

\usepackage{amsmath,bm,amsfonts,amssymb,array,calc,amsthm,rotating}
\usepackage[nosort]{cite} 
\usepackage{graphicx}
\usepackage{color}
\usepackage[colorlinks=true]{hyperref}

\renewcommand\baselinestretch{1.25} 
\renewcommand\section{\@startsection {section}{1}{\z@}%
                                   {-3.5ex \@plus -1ex \@minus -.2ex}%
                                   {2.3ex \@plus.2ex}%
                                   {\normalfont\large\bfseries}} 
\renewcommand\subsection{\@startsection{subsection}{2}{\z@}%
                                   {-3.25ex\@plus -1ex \@minus -.2ex}%
                                   {1.5ex \@plus .2ex}%
                                   {\normalfont\normalsize\bfseries}} 
\renewcommand\subsubsection{\@startsection{subsubsection}{3}{\z@}%
                                   {-3.25ex\@plus -1ex \@minus -.2ex}%
                                   {1.5ex \@plus .2ex}%
                                   {\normalfont\normalsize\it}} 
\renewcommand\paragraph{\@startsection{paragraph}{4}{\z@}%
                                   {-3.25ex\@plus -1ex \@minus -.2ex}%
                                   {1.5ex \@plus .2ex}%
                                   {\normalfont\normalsize\bf}} 
\renewcommand\subparagraph{\@startsection{subparagraph}{5}{\z@}%
                                   {-1.25ex\@plus -1ex \@minus -.2ex}%
                                   {0ex \@plus .2ex}%
                                   {\normalfont\normalsize\it}}

\numberwithin{equation}{section}

\long\def\@makecaption#1#2{%
  \vskip\abovecaptionskip
  \sbox\@tempboxa{{\bf #1:} #2}%
  \ifdim \wd\@tempboxa >\hsize
    {\small\bf #1:} {\small #2}\par
  \else
    \global \@minipagefalse
    \hb@xt@\hsize{\hfil\box\@tempboxa\hfil}%
  \fi
  \vskip\belowcaptionskip}

\renewcommand*\l@section[2]{%
  \ifnum \c@tocdepth >\z@
    \addpenalty\@secpenalty
    \addvspace{.5em \@plus\p@}%
    \setlength\@tempdima{1.5em}%
    \begingroup
      \parindent \z@ \rightskip \@pnumwidth
      \parfillskip -\@pnumwidth
      \leavevmode \bfseries
      \advance\leftskip\@tempdima
      \hskip -\leftskip
      #1\nobreak\hfil \nobreak\hb@xt@\@pnumwidth{\hss #2}\par
    \endgroup
  \fi}
\renewcommand*\l@subsection{\addvspace{.0em \@plus\p@}\@dottedtocline{2}{1.5em}{2.3em}}
\renewcommand*\l@subsubsection{\addvspace{-.2em \@plus\p@}\@dottedtocline{3}{3.8em}{3.2em}}

\def\hepth#1{\href{http://xxx.arxiv.org/abs/hep-th/#1}{{arXiv:hep-th/#1}}}

\def\mathsg#1{\href{http://xxx.arxiv.org/abs/math.SG/#1}{{arXiv:math.sg/#1}}}
\def\mathag#1{\href{http://xxx.arxiv.org/abs/math.AG/#1}{{arXiv:math.ag/#1}}}
\def\alggeom#1{\href{http://xxx.arxiv.org/abs/alg-geom/#1}{{arXiv:alg-geom/#1}}}

\definecolor{refcol}{rgb}{0.2,0.2,0.8}
\definecolor{eqcol}{rgb}{.6,0,0}
\definecolor{purple}{cmyk}{0,1,0,0}

\gdef\@citecolor{refcol}
\gdef\@linkcolor{eqcol}
\def\colorlinkspurple{\gdef\@urlcolor{purple}}
\def\colorlinksblue{\gdef\@urlcolor{blue}}
\def\colorlinksred{\gdef\@urlcolor{red}}

\def\eg{{\it e.g.}} 
\def\etc{{\it etc.}}

\def\revise#1       {\raisebox{-0em}{\rule{3pt}{1em}}%
                     \marginpar{\raisebox{.5em}{\vrule width3pt\ 
                     \vrule width0pt height 0pt depth0.5em 
                     \hbox to 0cm{\hspace{0cm}{%
                     \parbox[t]{4em}{\raggedright\footnotesize{#1}}}\hss}}}}

\def\cale         {{\cal E}} 
\def\calf         {{\cal F}}

\def\call         {{\cal L}} 
\def\calm         {{\cal M}} 
\def\caln         {{\cal N}} 
\def\calo         {{\cal O}}

\def\calt         {{\cal T}}

\def\calw         {{\cal W}} 

\def\complex      {{\mathbb C}} 
 
\def\projective   {{\mathbb P}} 
 
\def\reals        {{\mathbb R}} 
\def\zet          {{\mathbb Z}} 

\def\del          {\partial} 
\def\delbar       {\bar\partial} 
\def\ee           {{\it e}} 
\def\ii           {{\it i}} 
 
\def\tr           {{\rm Tr}}

\newcommand\topa[2]{\genfrac{}{}{0pt}{2}{\scriptstyle #1}{\scriptstyle #2}}

\def\sqr#1#2{{\vcenter{\vbox{\hrule height.#2pt   
 \hbox{\vrule width.#2pt height#1pt \kern#1pt 
 \vrule width.#2pt}\hrule height.#2pt}}}}

\def\CP{\complex\projective}
\def\RP{\reals\projective}

\def\eps{\epsilon}

\def\Ip{\mathord{\mathchar "0271}}
\def\Ipp{\mathord{\mathchar "0271 \kern-4.5pt \mathchar"0271}}

\def\TT{{\mathbb T}}
\def\aut{\mathop{\rm Aut}}

\def\eul{{\bf e}}

\catcode`\@=12 

\begin{document} 


\title{Opening Mirror Symmetry on the Quintic}

\pubnum{%
hep-th/0605162}
\date{May 2006}

\author{
Johannes Walcher \\[0.2cm]
\it School of Natural Sciences, Institute for Advanced Study\\
\it Princeton, New Jersey, USA
}

\Abstract{
Aided by mirror symmetry, we determine the number of holomorphic disks
ending on the real Lagrangian in the quintic threefold. The tension of 
the domainwall between the two vacua on the brane, which is the 
generating function for the open Gromov-Witten invariants, satisfies 
a certain extension of the Picard-Fuchs differential equation governing 
periods of the mirror quintic. We verify consistency of the
monodromies under analytic continuation of the superpotential 
over the entire moduli space. We reproduce the first few instanton 
numbers by a localization computation directly in the A-model, and 
check Ooguri-Vafa integrality. This is the first exact result 
on open string mirror symmetry for a compact Calabi-Yau manifold. 
}

\makepapertitle

\body

\version\versionno

\vskip 1em


\section{Introduction and Summary}

It has long been suspected that the enumerative results about
holomorphic curves obtained by mirror symmetry \cite{cdgp} could be 
extended to open Riemann surfaces, provided appropriate boundary 
conditions are imposed. In the A-model, and at lowest order in the string 
coupling expansion, the counting of holomorphic disks ending on
Lagrangian submanifolds is the central ingredient in the definition of
Floer homology and the Fukaya category \cite{fooo}, which appears
on one side of the homological mirror symmetry conjecture
\cite{icm}. From the physics perspective, the chief interest is 
to determine the superpotential on the worldvolume of D-branes 
wrapping the Lagrangian, with many applications in studies of $\caln=1$ 
compactifications of string theory.

Until now, the program of extending mirror symmetry to the open string
sector has been successfully implemented only in a rather limited set of
examples with special, toric, symmetries \cite{av1,akv}. While certain
general structures could be extracted from the results obtained 
\cite{mayr,lmw,lmw2}, and of course much is known in lower-dimensional
situations \cite{poza,bhlw}, it has remained unclear whether and how these
ideas could be implemented for more general, in particular compact, 
Calabi-Yau threefolds. This is precisely what we do in this paper.

The Calabi-Yau manifold $X$ we will consider is the most popular
quintic in $\CP^4$, and our Lagrangian $L$ will be the most 
canonical real locus inside of it. This Calabi-Yau-Lagrangian pair 
has been contemplated many times in the literature, starting with 
\cite{wcs}. First exact results were obtained in \cite{bdlr},
where D-branes wrapping $L$ where identified with certain
RS boundary states at the Gepner point \cite{resc} (see also
\cite{bhhw} for a complementary derivation of this result). In
\cite{howa,strings}, the continuation of these boundary states
over the moduli space was analyzed using matrix factorizations
\cite{kali1,bhls} in the mirror B-model Landau-Ginzburg description.
In particular, it was explained in \cite{strings} that the 
singularity in the D-brane moduli space at the Gepner point 
could be interpreted as a degeneration of the Morse-Witten-Floer 
complex that computes Floer homology. Living in the A-model, the
Floer differential differs from the classical Morse differential
by corrections from holomorphic disks ending on $L$ \cite{horietal,fooo},
which suggested that one should be able to turn these results
into a computation of the number of holomorphic disks as coefficients
in the appropriate large-volume expansion. We will fulfill this
promise in the present work, although following a slightly different
route.

The central technical discovery is that the spacetime superpotential 
on the brane worldvolume, which is the generating function capturing
the open string instanton information \cite{extending,oova,kklm},
satisfies a differential equation which is a simple extension of
the standard Picard-Fuchs differential equation whose solutions
are the periods of the holomorphic three-form on the mirror of the
quintic. The possible origin of such differential equations is 
discussed in special circumstances in \cite{lmw,lmw2} (see also 
\cite{indians,lema}). But for a general brane configuration, or when 
the ambient Calabi-Yau is compact, the existence of this differential
equation is, to the very least, surprising. Perhaps the most novel 
aspect of the equation that we introduce in this paper is that large 
complex structure is not a singular point of maximal unipotent 
monodromy. However, this has excellent reasons for being so, 
as we will explain below.

\begin{table}[t]
\begin{tabular}{|l|l|l|}
\hline
$d$ & number of disks $n_d$            & number of spheres \\\hline
1   & 30                               & 2875 \\
3   & 1530                             & 317206375 \\
5   & 1088250                          & 229305888887625  \\
7   & 975996780                        & 295091050570845659250 \\
9   & 1073087762700                    & 503840510416985243645106250 \\
11  & 1329027103924410                 & 1017913203569692432490203659468875 \\
13  & 1781966623841748930              & 229948856813626664832516010477226554$\ldots$ \\
15  & 2528247216911976589500           & 562465682466848327417948393837157975$\ldots$ \\
17  & 3742056692258356444651980        & 146020747145890338745688881159596996$\ldots$ \\
19  & 5723452081398475208950800270     & 397016669854518762338361058844977288$\ldots$ \\
\hline
\end{tabular}
\caption{The number (integral invariants) of holomorphic disks in 
$X$ ending on $L$, of degree $d$ (only odd $d$ are shown, for 
reasons explained in the text), and, for comparison, the number 
of holomorphic spheres in $X$, according to \cite{cdgp}.}
\label{open}
\end{table}

Equipped with the differential equation, it is
straightforward to extract the open string instanton numbers, 
and we can check the integrality property conjectured in \cite{oova}.
We do all this in section \ref{supo}, and display, for amusement, the 
results in table \ref{open}. 

It is then also of interest to study the analytic properties of 
the brane superpotential over the entire Calabi-Yau moduli space, 
and not just around large volume. Referring to section 
\ref{analytic} for details, we would like to point out two salient 
features here. Firstly, the domainwall tension is invariant under 
monodromy around the conifold singularity in the moduli space. To 
appreciate the consistency of this result, one has to remember that 
the cycle that shrinks to zero volume at the conifold singularity 
in K\"ahler moduli space is a holomorphic cycle which can be 
wrapped by a B-brane, and it would be somewhat non-obvious why 
an A-brane would feel this singularity. 

The second interesting feature is that the domainwall tension is 
not invariant under monodromy around the small-volume Gepner point. 
This is more surprising because, based on the worldsheet results of 
\cite{strings}, one would have naively expected the domainwall tension 
to vanish at that point where the two vacua on the brane become
degenerate. Instead, what happens is that the tension of the 
domainwall, when analytically continued from large volume, becomes 
asymptotically equal to a particular closed string period, which 
measures flux superpotentials. In other words, this domainwall 
only mediates a transition between different flux sectors, and 
this is still consistent with the degeneracy of the open string 
vacua. What it tells us, however, is that it could be much more
delicate to understand our results from the worldsheet perspective, 
which is purportedly insensitive to the flux. It also indicates
that it might be appropriate to include some of the flux data into 
the definition of Floer homology and the Fukaya category.

While this derivation of the superpotential and the instanton series
can perfectly well stand alone, the confidence in the enumerative 
results of table \ref{open} of course increases dramatically if at
least some of those numbers can be verified mathematically directly
in the A-model. We will do this in section \ref{localization}.

The mathematical definition of open Gromov-Witten invariants in
general still appears lacking \cite{fooo}, although several special
cases have been treated in the literature. Studies of the local 
toric situation include \cite{katzliu,grza,peter,liu}. Recently, 
Solomon \cite{jakethesis,jakecolumbia} has performed a rigorous 
study of open Gromov-Witten invariants in the situation in which 
the Lagrangian providing the boundary conditions arises as the 
fixed point set of an anti-holomorphic involution. This covers 
the situation of our interest, so we can be confident that the 
numbers we are claiming are well-defined.

To go ahead with the direct computation of those open Gromov-Witten
invariants, one can exploit the fact that, at least in our situation,
any holomorphic mapping from the disk into $X$ with boundary on $L$ 
factors through a holomorphic sphere in $X$ meeting $L$ in a circle. 
In other words, we can relate the enumeration of holomorphic disks
to the enumeration of holomorphic spheres which are invariant under
the anti-holomorphic involution. For this problem, we have at our
disposal the powerful graph combinatorial method introduced
in \cite{kontsevich}. This technique computes the Euler characteristic
of a particular bundle on the moduli space $\calm(\CP^4)$ of 
holomorphic curves in $\CP^4$ by using Atiyah-Bott localization 
with respect to the action of the torus $(S^1)^5\subset U(5)$ inside 
the symmetry group of $\CP^4$. The anti-holomorphic involution then 
acts in a natural way on this moduli space and the bundle over it, 
and one can identify the open Gromov-Witten invariant as the Euler 
characteristic of the resulting real bundle over the real locus 
in $\calm(\CP^4)$ \cite{jake}.

There are then two key points to appreciate in order to proceed. The 
first one is that while the anti-holomorphic involution breaks 
some of the symmetries of the ambient space, it still leaves an 
$(S^1)^2\subset O(5)$ unbroken. In particular, the fixed points on
the real slice with respect to this torus coincide with the real 
fixed points of the torus in the complex case. The second point is 
that the Euler class of a real bundle is the squareroot of the 
Euler class of its complexification, where the sign is determined 
by the choice of orientation. With these two ingredients, it is
straightforward to adapt the methods of \cite{kontsevich} to
develop a graphical calculus which computes the open Gromov-Witten
invariants of our interest. We have checked that up to degree 7,
these numbers coincide with those obtained using mirror symmetry.
The number (30) of holomorphic disks of degree 1 was first computed
(without using localization) by Solomon \cite{jake,jakecolumbia}. 
We have also checked the number (1530) of holomorphic disks in 
degree $3$ by taking a real slice of the localization computation 
of \cite{es} on the space of curves (instead of the space of maps).

Besides the many possible applications and extensions of these 
results that spring to mind, we would like to mention that the 
numbers we get in this paper can also be viewed as providing
lower bounds in real enumerative geometry in the sense of, see,
\eg, \cite{sottile,welschinger}.

\section{The problem and its solution}
\label{supo}

We consider in $\CP^4$ the Calabi-Yau hypersurface given as the vanishing
locus of a polynomial of degree $5$ in the homogeneous coordinates of
$\CP^4$:
\begin{equation}
X = \{ P(z_1,\ldots,z_5) = 0 \} \subset \CP^4
\end{equation}
The choice of $P$ determines the complex structure of $X$, and to define 
a $\sigma$-model with target space $X$, we need to pick a choice of
complexified K\"ahler form $B+\ii J = t\omega$, where we denote by 
$\omega$ the integral generator of $H^2(X,\zet)=\zet$, and $t$ is
the K\"ahler parameter.

\subsection{On the real quintic}

We want to identify in $X$ a particular Lagrangian submanifold as the
fixed-point locus of an anti-holomorphic involution which 
acts on the ambient $\CP^4$ as complex conjugation on the homogeneous
coordinates
\begin{equation}
\eqlabel{conjugate}
[z_1:z_2:\cdots:z_5] \mapsto [\bar z_1:\bar z_2:\cdots:\bar z_5]
\end{equation}
The complex structure on $X$ will be (anti-)invariant under this involution
if the defining polynomial $P$ is {\it real}, in the sense that all its 
coefficients are real (up to a common phase). The fixed point locus, $L$,
where $z_i=x_i$ is real is then given by the corresponding real equation 
$P(x_1,\ldots,x_5)=0$ inside of $\RP^4\subset \CP^4$. Straightforwardly, 
$L$ is a Lagrangian submanifold of $X$. In fact, $L$ is even special 
Lagrangian with respect to the holomorphic three-form on $X$.

Now while the topology of $X$ is well-known and independent of the complex
structure, the real locus $L$ can have various topologies and singularities,
with interesting transitions between them as $P$ is varied. We will not 
attempt to discuss all the possibilities here, but wish to comment on the
consequences. To fix ideas, let us consider the Fermat quintic
\begin{equation}
\eqlabel{fermat}
P = z_1^5+z_2^5+z_3^5+z_4^5+z_5^5
\end{equation} 
Over the reals, $z_i=x_i$, we can solve for $x_5$ uniquely in terms of 
$x_1, \ldots x_4$, not all of which can be zero, lest $x_5$ will be zero 
too. This identifies $L$ with a copy of $\RP^3$. However, this 
identification depends on the fact that $z_5^5=a$ for real $a$ has 
only one real root, which will not be useful for a generic $P$.

There are at least two things that can happen to the real locus as 
we vary the complex structure. The first one is familiar from studies
of stability conditions on Lagrangian submanifolds, and happens along
a real codimension one locus in complex structure moduli space. When
crossing such a wall of marginal stability, the special Lagrangian $L$ 
develops a singularity and reconnects on the other side, changing its
topological type (but not its homology class). The second effect is 
a remnant of the standard conifold singularity in the complex structure
moduli space. (It might seem that since the discriminant locus is 
complex codimension one, it would generically be missed by the 
half-dimensional real subspace. But this is untrue.) It was shown 
in \cite{hhprw} using a local model that when crossing such a conifold 
singularity, the homology class of the real locus always changes by 
the homology class of the vanishing cycle.

The second phenomenon is known to happen on the quintic \cite{unpublished}, 
for example when crossing the standard conifold locus $\psi=1$ along the 
one parameter family $P\to P-5\psi z_1z_2z_3z_4z_5$. Since the Lagrangian
is connected at $\psi=0$, it implies that we must also be crossing a line
of marginal stability somewhere between $\psi=0$ and $\psi=\infty$.

In this paper, we are studying $L$ in the A-model, and those aspects 
should be independent of the complex structure of $X$, and only depend
on the Hamiltonian deformation class of $L$. Namely, we would
expect to only depend on $L$ being Lagrangian, and not the special
Lagrangian property. On the other hand, the available definitions 
of Floer homology for Lagrangians clearly depend on the underlying 
topology. (For instance, they depend on $b_1(L)$.) It is therefore not 
a priori clear why there should be a well-defined and invariant notion 
of Floer homology or of the ``number of disks'' ending on ``the real locus
$L$'' which is independent of the complex structure of $X$. One might 
worry slightly less about this in regard to the first phenomenon (marginal 
stability) because at least the homology class is preserved. In this paper, 
in any case, we will ignore this complication, and just pretend that 
$L\cong \RP^3$. The number of disks we will quote can then be understood 
as referring to ``the generic quintic in a neighborhood of the Fermat 
point''. 

For the rest of the paper, we will be concerned with the dependence 
on the K\"ahler parameter, $t$, or its exponentiated version 
$q=\ee^{2\pi \ii t}$. We begin in the large volume limit $q\to 0$.

\subsection{Vacuum structure at large volume}
\label{vacua}

Recall that to wrap an A-brane on $L$, we also need to specify a 
$U(1)$ bundle with a flat connection. Since $H_1(L;\zet)=\pi_1(L)=
\zet_2$ we have two possible choices which are distinguished by
a ``discrete Wilson line'', $W=\epsilon=\pm1$. In fact, these two 
choices correspond to topologically distinct bundles on $\RP^3$, as 
measured by the first Chern class $c_1\in H^2(L;\zet)$. The latter is 
equal to $H_1(L;\zet)$ by Poincar\'e duality. On the other hand, the 
K-theory of the quintic does not contain any torsion elements, and
the two choices of flat connection can therefore not be distinguished 
by any topological charge \cite{bdlr}.

As a consequence, when wrapping a D6-brane of type IIA string theory
on $L$, the brane worldvolume will support an $\caln=1$ gauge theory 
with two vacua corresponding to the two possible discrete Wilson
lines, which are not distinguished by any conserved charge. We can then
ask about the existence of a BPS domainwall that communicates between
these two vacua. 

To represent this domainwall in string theory, it is helpful to 
understand why the two bundles are topologically equivalent
after inclusion in the quintic. Let us consider the situation with
``non-trivial'' Wilson line $\eps=-$ (we will see in a moment that this
isn't really an invariant notion). The non-trivial first Chern class 
of the bundle on $L$ can be viewed as resulting from dissolving into 
the D6-brane a D4-brane wrapping the non-trivial one-cycle in 
$H_1(L;\zet)$. But since the quintic does not contain any non-trivial 
one-cycles, we can also contract it away to nothing. 

Clearly, then, the BPS domainwall that mediates between the two choices 
of Wilson line on the D6-brane wrapping on $L$ is a D4-brane wrapping 
a holomorphic disk $D$ in $X$ with boundary on the non-trivial one-cycle 
in $L$ and extended along a (2+1)-dimensional subspace of Minkowski
space. This D4-brane is a magnetic source on the D6-brane and hence
changes the (discrete) magnetic flux on $L$. The topological 
classification of $D$ is as a non-trivial relative cohomology class
in $H_2(X,L;\zet)$ with non-trivial image in $H_1(L;\zet)$.

It is not difficult to get a first approximation to the tension, $\calt$, of 
this domainwall in the large volume limit (here and throughout the paper, 
we will refer to the tension as the holomorphic quantity whose absolute 
value gives the physical tension). Since $L$ is defined as the fixed 
point locus of an anti-holomorphic involution of $X$, any holomorphic 
disk ending on $L$ can be complex conjugated to a second holomorphic 
disk, and thereby completed to a holomorphic sphere. From the exact 
sequence
\begin{equation}
\eqlabel{exact}
H_2(X;\zet) \to H_2(X,L;\zet) \to H_1(L;\zet)
\end{equation}
we see that in fact also a brane wrapped on twice the generator 
of $H_2(X,L;\zet)$ will not change the vacuum on the brane, and hence 
be equivalent to a holomorphic sphere. The tension of that sphere being 
$t$ (the K\"ahler parameter), we infer $2\calt\sim t$.

To see that this argument was in fact quite incomplete, we need another 
fact about the relation between the cohomology of $X$ and that of $L$.
Namely, when intersecting a hyperplane in $\CP^4$ with the Lagrangian $L$
(the hyperplane has to be represented by a complex linear equation in
order to intersect $L$ transversely), we can see that the intersection
locus is a non-trivial one-cycle in $L$. The Poincar\'e dual statement 
is that the integral generator of $H^2(X;\zet)$ restricts on $L$ to the 
non-trivial element of $H^2(L;\zet)$. Since the gauge invariant gauge
field on the brane is $B-F$, this means that changing the flat $B$-field 
on $X$ by one unit is equivalent to exchanging the two flat gauge 
fields on the brane.

A more elementary way to see this is to note that the path-integral
contribution of a disk worldsheet wrapped on $D$ has a contribution 
$\ee^{2\pi\ii t/2}=q^{1/2}$ from its area and a contribution $\eps=\pm 1$ 
from its boundary, so changing $B\to B+1$ is equivalent to changing
$\eps\to -\eps$. Taking $B\to B+2$ does nothing on the brane. In this sense,
we can specify the Wilson line on the brane only after fixing the
sign of $q^{1/2}$.

Now claiming that $\calt\sim t/2$ raises a puzzle because it is not 
invariant under $t\to t+2$. To resolve this, we have to note that
the D4-brane wrapped on $D$ is a magnetic source not only for the
gauge field on $L$, but also for the Ramond-Ramond 3-form field
(we actually used this above to derive $2\calt\sim t$). The change of
$\calt$ under $t\to t+2$ is then explained by the non-invariance of
RR flux under $B$-field monodromies.

So to make the formula for $\calt$ more precise, and work out the spectrum
of domainwalls, we have to include the RR flux quantum numbers in our 
labeling of the vacua. For the time being, 4-form flux, $N_4$, and 
6-form flux, $N_6$, (around the unique four and 6-cycle of $X$) will 
suffice, so our vacua are labeled as $(N_4,N_6,\eps)$. 

We then require that a domainwall represented by a D4 wrapping an 
elementary disk $D$ connects $\eps$  $-\eps$, and that by juxtaposing 
two such disks we obtain a sphere across which the only change is 
$N_4\to N_4+1$, that the B-field monodromy $B\to B+1$ changes $N_6
\to N_6+N_4$, and also $\eps\to -\eps$, but is otherwise a symmetry 
of the spectrum. We also wish to keep 4- and 6-form flux integrally 
quantized to avoid concluding with fractional D0-branes. 

It then turns out that, up to parity, there is only one consistent 
solution to these constraints. The change in 4-form flux across a 
D4-brane wrapped on $D$ is zero when $\eps=-$ on the left of the 
domainwall and it is equal to $+1$ when $\eps=+$ on the left, and,
we have to let the $B$-field monodromy {\it change the 4-form flux}, 
in a way depending on $\eps$:
\begin{equation}
\eqlabel{spectrum}
B\to B+1:\qquad
\begin{array}{rcl}
(N_4,N_6,-)&\to& (N_4,N_6+N_4,+) \\
(N_4,N_6,+)&\to& (N_4+1,N_6+N_4,-)
\end{array}\qquad\quad
\end{equation}
Let us denote the tension of a domainwall between vacuum $(N_4,N_6,\eps)$ 
on the left and vacuum $(N_4',N_6',\eps')$ on the right by 
$\calt_{(N_4,N_6,\eps)|(N_4',N_6',\eps')}$. The above constraints are enough 
to determine all $\calt$'s as a function of $t$.

For example, let us consider the most basic $\calt_- \equiv 
\calt_{(0,0,-)|(0,0,+)}$ and $\calt_+\equiv \calt_{(0,0,+)|(1,0,-)}$. Since $\calt_-(t+1) 
= \calt_+(t)$ and $\calt_++\calt_-=t$, we conclude
\begin{equation}
\eqlabel{classten}
\calt_- = \frac t2-\frac14 \qquad\qquad
\calt_+ = \frac t2+\frac14
\end{equation}
Finally, we can write down the spacetime superpotential, which follows 
from \eqref{classten} together with
\begin{equation}
\calt_{(N_4,N_6,\eps)|(N_4',N_6',\eps')}(t) =
\calw_{N_4',N_6',\eps'}(t) - \calw_{N_4,N_6,\eps}(t)
\end{equation}
We find
\begin{equation}
\eqlabel{classsupo}
\calw_{N_4,N_6,+}(t) = \frac {t^2}4 + N_4 t + N_6 
\qquad\quad
\calw_{N_4,N_6,-}(t) = \frac{t^2}4 - \frac t2+\frac 14 + N_4 t + N_6
\end{equation}
Of course, in this section, the discussion has been entirely classical
and restricted to the large volume limit $t\to\ii\infty$. We now proceed
to study the corrections $\calw^{\rm quant.}$ from worldsheet instantons.

\subsection{Worldsheet instanton corrections}

According to general philosophy \cite{kklm,oova,extending,av1,bcov}, the
spacetime superpotential on the worldvolume of a particular supersymmetric
brane wrapping a cycle in a Calabi-Yau manifold, $X$, when expressed in the 
A-model, and expanded in the appropriate variables, becomes the generating 
function counting worldsheet instanton corrections from holomorphic disks 
ending on the Lagrangian, $L$. Such a statement is in line with the role that 
holomorphic disks play in the definition of Fukaya's $A_\infty$ 
category \cite{fooo}, and the relationship between $A_\infty$ algebras 
and D-brane superpotentials \cite{calin,tomasiello}.

More precisely, the spacetime superpotential can be identified with the
topological disk partition function and is conjectured to admit an
expansion of the general form
\begin{equation}
\eqlabel{expansion}
\calw(t,u)  = F_{\rm disk} (t,u)
= \sum_{d,e} {\tilde n}_{d,e} q^d y^e = \sum_{d,e}\sum_{k\ge 1}
 \frac{n_{d,e}}{k^2} q^{kd} y^{ke}
\end{equation}
Here, the sum is over relative cohomology classes in $H_2(X,L)$, $q=\ee^{2\pi\ii t}$ 
is the (collection of) closed string K\"ahler parameters of $X$ and 
$y=\ee^{2\pi\ii u}$ is the (collection of) exponentiated classical open string 
deformation parameters. The latter come from non-Hamiltonian deformations 
of the Lagrangian. They are $b_1(L)$ in number and are complexified by the
Wilson line of the gauge field around the corresponding one-cycles
of $L$. The final transformation in \eqref{expansion} is a resummation 
of multi-cover contributions and the central part of the conjecture
is that the resulting expansion coefficients $n_{d,e}$ are integers 
\cite{oova} (whereas the $\tilde n_{d,e}$ are in general rational
numbers). These integers have a spacetime interpretation as counting 
the ``degeneracy of BPS domainwalls'' in the class $(d,e)$.

The existence and integrality of such an expansion has been checked in 
many examples involving local toric Calabi-Yau manifolds. Our goal in 
this paper is to make sense of and evaluate the formula \eqref{expansion} 
for the Calabi-Yau-Lagrangian pair $(X,L)= \text{(quintic, real locus)}$.
At first sight, the fact that we only have a discrete open string 
modulus at our disposal is a deficiency because \eqref{expansion} 
makes explicit only rational cohomology. On second thought, however, 
it's a blessing. 

For example, as we have discussed above, domainwalls arising from 
D4-branes wrapping holomorphic disks are sources for both the 
Ramond-Ramond field and the gauge field on the brane. But if the 
disk ends in a rational cycle of $L$, the gauge flux is non-zero
as a differential form. This raises a puzzle because according to 
the standard worldsheet analysis, gauge fields on Lagrangian A-branes 
should be flat. From the spacetime perspective, this might well be 
repaired by a careful analysis of the couplings of the brane to the 
Ramond-Ramond fields. But it is clearly not obvious to see that from 
the TFT on the worldsheet. In the cases discussed in the literature
(see \cite{av1,akv} and follow-up work), this problem is avoided
because the Lagrangians considered there are non-compact and hence 
the flux can disperse to infinity.

A second advantage of having $H_1(L,\zet)=\zet_2$ being torsion
has to do with certain puzzlements \cite{jake} about the multi-cover 
formula as well as the integral ``framing'' ambiguity of open string 
amplitudes discovered in \cite{akv}. We do not understand either of 
those issues sufficiently well enough to usefully discuss here, but 
the consistency of our results indicates that both problems are 
absent for $H_1(L)=\zet_2$.

Finally, because our Lagrangian is compact, we can also discuss
the classical contributions to the superpotential, as we have
done in the previous subsection. The structure of these classical
terms (which are absent from \eqref{expansion}) will help us to
normalize the computation by imposing consistency of the monodromies
around the various singular loci in the K\"ahler moduli space (see
section \ref{analytic}).

So what is the possible structure of worldsheet instanton corrections
to our formulas \eqref{classten} for the domainwall tensions?

Clearly, the first non-trivial term will arise from worldsheet disks
wrapped in the class $D$ generating $H_2(X,L;\zet)=\zet$, and will 
contribute at order $q^{1/2}$. Then there will be higher order terms.
Let us call disks contributing at order $q^{d/2}$ ``of degree $d$''.
It is easy to see that the conditions $\calt_-(t+1)=\calt_+$, $\calt_++\calt_-=t$
that we have used to derive \eqref{classten} hold also after inclusion
of non-perturbative worldsheet corrections. This is because
$t$ is essentially {\it defined} to be the parameter measuring the
tension of the domainwall wrapped on a degree $1$ rational curve.
The only form of the instanton expansion that is consistent with those
constraints is that there are no contributions from even degree disks.
This is in fact not unexpected, because disks of even degree have 
trivial boundary on the Lagrangian, and even though we can contemplate
holomorphic disks of even degree ending on $L$, the triviality of their
boundary makes it difficult to keep them there as we vary the complex
structure of the quintic. In other words, we do not expect any invariant 
to exist for even degree. 

So we expect a result of the form
\begin{equation}
\eqlabel{expected}
\calt_\pm = \frac t4 \pm \frac 12 \pm {\it const.} \sum_{d\; {\rm odd}} \tilde n_d q^{d/2}
\end{equation}
where the $\tilde n_d$ are certain rational numbers such that rewriting
them as in \eqref{expansion},
\begin{equation}
\tilde n_d = \sum_{k|d} \frac{n_{d/k}}{k^2}
\end{equation}
the $n_d$ turn out integer.

\subsection{Mirror Symmetry and open Picard-Fuchs equation}

The easiest way to get an expansion of the form \eqref{expected} is to
make use of mirror symmetry. What this means concretely is that we
should first identify an object in the D-brane category which appears on 
the B-model side of the homological mirror symmetry conjecture, and
which, via the equivalence of categories and up to auto-equivalences,
corresponds to the object of the (derived) Fukaya category that is
defined by $L$. We should then compute the appropriate 
superpotential/domainwall tension quantity as a function of the mirror
parameter $\psi$ and reexpress it in terms of the flat coordinate $t$.

The Calabi-Yau mirror, $Y$, to the quintic is of course well-known. It 
is the resolution of a $(\zet_5)^3$ quotient of the one-parameter family 
of quintics $\sum z_i^5-5\psi\prod z_i=0$ in $\CP^4$. Equivalently,
we can consider a Landau-Ginzburg orbifold model with superpotential
$W=\sum z_i^5-5\psi \prod z_i$ and orbifold group $(\zet_5)^4$. The
corresponding B-model category which is conjectured \cite{stability} 
to be equivalent to the derived category of $Y$ is the category 
of $(\zet_5)^4$ equivariant matrix factorizations of the superpotential 
$W$. (The corresponding equivalence was proven for the quintic itself
by Orlov \cite{orlov}.) 

And in fact, as we have mentioned in the introduction, the matrix 
factorization which is mirror to the Lagrangian $L$ is known explicitly
(see \cite{howa,strings} for details). Given this identification of 
the matrix factorization and the equivalence with the derived category, 
it should be possible in principle to also describe explicitly a coherent 
sheaf on $Y$ corresponding to $L$. This would in fact be very interesting, 
because it would allow making use of some of the well-known machinery of 
holomorphic vector bundles that applies to problems of this type. 
In particular, there is an explicit formula for the superpotential, 
namely, the holomorphic Chern-Simons functional \cite{wcs}
\begin{equation}
\eqlabel{hcs}
\calw^B = S_{\rm hCS}(A,A_0) = \int\Omega\wedge \tr \bigl[ A\wedge 
\delbar_{A_0} A + \frac 23 A\wedge A\wedge A\bigr]
\end{equation}
No such expression is known in the matrix factorization formulation, 
and although there are formulas for TFT correlators \cite{kali2,hela},
they do not appear sufficient to determine the full superpotential.
(See, however \cite{hln} for recent progress in making the 
$A_\infty$ constraints of \cite{hll} useful for this type of question.)

Leaving these explicit B-models for future investigations, we will instead
obtain sufficient guidance from the non-compact examples of open mirror 
symmetry introduced in \cite{av1}, and studied in depth in 
\cite{akv,mayr,indians,lema,lmw,lmw2}.

The main simplification that occurs in these examples is that the B-model
contains only D5-branes wrapped on curves in the Calabi-Yau. For such 
a brane configuration, the holomorphic Chern-Simons action \eqref{hcs}
reduces to a ``partial period'' integral of the type
\begin{equation}
\eqlabel{mina}
\calw(C,C_*) = \int_\gamma \Omega
\end{equation}
where $\gamma$ is a three-chain in $X$ with boundary $\del\gamma=C-C_*$
equal to the difference of two possible positions of the D5-branes.
(If $C$ and $C_*$ are holomorphic, \eqref{mina} is literally the tension
of the domainwall between the two vacua.) In the toric case, one can
then further reduce the integral \eqref{mina} to take place on a
Riemann surface, so one has essentially a one-dimensional problem.

This structure was exploited in \cite{lmw,lmw2} to show that the 
differential equations obtained in \cite{mayr,lema} could be viewed 
as resulting from a certain variation of mixed Hodge structure on 
a certain relative cohomology. Explicitly, one retains the boundary
terms arising in the derivation of the GKZ differential system and
converts them into appropriate boundary variations. The upshot is 
that the open string mirror computations in the local toric case can 
be cast in a form very similar to the standard, closed string computations, 
involving Picard-Fuchs differential equations, maximal unipotent 
monodromy, mirror map, \etc. This is called $\caln=1$ special geometry.

We do not know at present whether such considerations make sense for the 
general B-model situation. The case at hand, however, is sufficiently well 
constrained by our results so far that assuming the existence of a 
differential equation with properties as in \cite{lmw,lmw2}, there 
is essentially a unique candidate. This moreover turns out to produce
excellent results.

The central idea of $\caln=1$ special geometry is to extend the standard
period vector by certain ``partial periods'' encoding information about
the open string sector. We recall that in standard ($\caln=2$) special 
geometry, we have two periods for every closed string modulus, plus one 
or two extra ones related to the holomorphic three-form. In $\caln=1$ 
special geometry, we gain one ``partial period'' for every classical 
open string modulus, plus one for every brane vacuum included in the 
background. Schematically,
\begin{equation}
\eqlabel{period}
\Pi(t_{\rm closed},u_{\rm open}) = (1, t_{\rm closed}, \del_t 
\calf_{\rm closed}, u_{\rm open}, \calw_{\rm brane}, \ldots )^T
\end{equation}
where $\calf_{\rm closed}$ is the standard prepotential and the 
$u_{\rm open}$ are the flat coordinates of the open string sector. The 
important point is that the period vector \eqref{period} satisfies a 
certain extension of the Picard-Fuchs differential equations. This 
differential system has all of the closed periods as solutions, plus 
extra ones related to $u_{\rm open}$ and $\calw_{\rm brane}$. The 
latter gives the open string instanton expansion according to 
\eqref{expansion}.

In the case that we have discussed in the previous subsections, we are
not adding any classical open string modulus because $b_1(\RP^3)=0$,
so the only modulus is the K\"ahler parameter $t$ of $X$, or equivalently,
the mirror variable, $z=z(t)$. Moreover, according to \eqref{expected}, 
we need exactly one non-trivial domainwall tension as function of $t$ 
to encode the desired open string expansion. Let us call 
$\tau \sim q^{1/2} + \cdots$ the quantum part of the expansion \eqref{expected}.
Since to leading order $z = q=\ee^{2\pi\ii t}$, we will also have $\tau(z)
\sim z^{1/2}+\cdots$ when expressed as a function of $z$.

Thus, we are simply seeking an ordinary linear differential equation in 
$z$, which, in addition to the four known periods of the mirror quintic, 
has exactly one additional linearly independent solution, $\tau$, with a 
squareroot behavior at $z=0$. The Picard-Fuchs equation governing periods 
of the mirror quintic being
\begin{equation}
\eqlabel{picard}
\call\varpi = \bigl[\theta^4 - 5 z (5\theta+1)(5\theta+2)(5\theta+3)(5\theta+4)
\bigr]\varpi = 0
\end{equation}
where $\theta = z\del_z$, and $z=(5\psi)^{-5}$, virtually the only possible
extension that satisfies our constraints is the differential operator
\begin{equation}
\eqlabel{opf}
(2\theta-1)\call = (2\theta-1)\theta^4 -
5 z(2\theta+1)(5\theta+1)(5\theta+2)(5\theta+3)(5\theta+4)
\end{equation}
We will now analyze this differential equation and show that it satisfies
all the other desirable properties as well.

\subsection{The instanton sum}

We follow conventions of \cite{cdgp}. The differential equation 
$\call\varpi=0$ has one distinguished solution, called the fundamental 
period, which has a power series expansion around the large complex 
structure point $z=0$,
\begin{equation}
\eqlabel{fundamental}
-w^2(z) \equiv \varpi_0(z) = \sum_{m=0}^\infty \frac{(5m)!}{(m!)^5} z^m
\end{equation}
All other solutions contain logarithms as $z\to 0$, large complex
structure being a point of maximal unipotent monodromy. The period
with a single logarithm, $w^1(z)$, has the information about
the mirror map via $t=w^1/w^2$, $q\equiv \ee^{2\pi\ii t}$.
\begin{equation}
\eqlabel{mirrormap}
- 2\pi\ii w^1(z) = \varpi_0(z) \log z + 5 \sum_{m=1}^\infty
\frac{5m)!}{(m!)^5} z^m\bigl[\Psi(1+5m)-\Psi(1+m)\bigr]
\end{equation}
Under large complex structure monodromy, $z\to \ee^{2\pi\ii} z$,
$w^1\to w^1+w^2$ and $t\to t+1$.

There are then two further solutions of \eqref{picard}, both of which
contain the closed string instanton information, in slightly different
forms. Specifically, the solution of \eqref{picard} called $\calf_1$ in
\cite{cdgp} is characterized by the boundary conditions
\begin{equation}
(2\pi\ii)^2 \calf_1 = -5\cdot (2\pi\ii) w^1(z) \log z + \frac 52 w^2(z)(\log z)^2 
-\frac{21}2\cdot (2\pi\ii)^2 w^1(z) + \calo(z)
\end{equation}
It transform under large complex structure monodromy as
$\calf_1\to \calf_1 - 5 w^1-8w^2$. Finally, the solution called 
$\calf_2$ in \cite{cdgp} is characterized by $\calf_2\to
\calf_2 -\calf_1 - 3 w^1 + 5w^2$ as $t\to t+1$. 

These periods $(\calf_1,\calf_2,w^1,w^2)$ can be interpreted as the quantum
corrected masses of D4, D6, D2 and D0-brane on the quintic, respectively 
\cite{bdlr}. They therefore also give the tension of domainwalls mediating
between various flux sectors, including the corrections from worldsheet 
instantons. For example, in the proper K\"ahler normalization $w^2=1$, one 
obtains after inverting \eqref{mirrormap} and expanding in $q=\ee^{2\pi\ii t}$,
\begin{equation}
\eqlabel{unconv}
\frac{\calf_1}{w^2} = - \frac 52 t^2 -\frac{21}2 t +\frac{1}{4\pi^2}\Bigl[
2875 q + \frac{4876875}4 q^2 +\cdots \Bigr]
\end{equation}
The polynomial in $t$ is the classical tension from the geometric volume of
the cycles and the power series in $q$ gives the quantum corrections. The
rational coefficient $\tilde N_d$ of $q^d$ in this expansion gives the
contribution from holomorphic spheres of degree $d$. They satisfy the
property that when reexpressed in terms of $N_d$ via
\begin{equation}
\tilde N_d = \sum_{k|d} \frac{d N_{d/k}}{k^3} \,,
\end{equation}
the $N_d$ are integers. Note that we have here slightly unconventionally
expanded the first derivative of the prepotential instead of the prepotential
itself or the Yukawa coupling as in \cite{cdgp}. Since periods and 
brane superpotentials are on equal footing in $\caln=1$ special geometry,
this will make the comparison with the open string version \eqref{opinst}
more natural.

Turning now to the equation \eqref{opf}, it has, by construction, exactly 
one additional solution, which we normalize to $\tau(z) = z^{1/2}+\cdots$. 
We find,
\begin{equation}
\eqlabel{tau}
\tau(z) = \frac{\Gamma(3/2)^5}{\Gamma(7/2)}\;
\sum_{m=0}^\infty \frac{\Gamma(5m+7/2)}{\Gamma(m+3/2)^5}\; z^{m+1/2}
\end{equation}
In the next section, we will determine from monodromy calculations on
the K\"ahler moduli space that $\tau$ enters the domainwall tension
in the normalization
\begin{equation}
\eqlabel{quantumten}
\calt_\pm (t) = \frac{w^1}{2} \pm \frac {w^2}{4} \pm \frac{15}{\pi^2} \tau(z)
\end{equation}
This then has exactly the expected form \eqref{expected}. Consulting 
\eqref{classten} and its relation with \eqref{classsupo}, we then conclude 
that the contribution of worldsheet disk instantons to the spacetime 
superpotential is
\begin{equation}
\calw^{\rm quant.} = \frac{30}{4\pi^2} \tau(z) 
\end{equation}
Dividing by $w^2$ to go to the canonical normalization of the holomorphic 
three-form, multiplying by $4\pi^2$ as in \eqref{unconv}, inverting the mirror 
map, and doing the expansion, we obtain the open string instanton sum
\begin{equation}
\eqlabel{opinst}
\hat\tau(q) = 30\frac{\tau(z(q))}{\varpi_0(z(q))} =
30 q^{1/2} + \frac{4600}3 q^{3/2} + \frac{5441256}5 q^{5/2} +\cdots
\end{equation}
We can then plug in to $\hat\tau(q)$ the Ooguri-Vafa multi-cover formula
\eqref{expansion}
\begin{equation}
\hat\tau(q) =  
\sum_{\topa{d\;{\rm odd}}{k\;{\rm odd}}} \frac{n_d}{k^2} q^{d k/2}
=
\sum_{d\;{\rm odd}} n_d \frac{q^{d/2}}{4} \Phi(q^d,2,1/2)
\end{equation}
where $\Phi$ is the Lerch Transcendent. For reasons explained in a previous
subsection, we only consider disks of odd degree and their odd multi-covers. 
The first few $n_d$ are indeed integer and displayed in table \ref{open} 
in the introduction.

It should be stressed that we have strictly speaking not shown that
the constant normalization factor in \eqref{expected} is equal to 
$\frac{1}{2\pi^2}$ as claimed. It is, however, the most natural choice
and consistent with everything else we know. It would be interesting
to derive this value more directly.

\section{Analytic continuation of the superpotential}
\label{analytic}

The purpose of this section is to analytically continue our result for
the superpotential/domainwall tension over the entire quantum K\"ahler
moduli space of the quintic, much as was done for the closed string 
periods in \cite{cdgp}. This will not only help us to fix the normalization 
factor anticipated in \eqref{quantumten}, but is interesting in its own 
right as it can shed light on intrinsically stringy aspects of D-brane
physics that have hitherto been inaccessible. We will indeed find that 
the analytic properties of the $\calt_\pm$ are rather interesting.

Recall that the K\"ahler moduli space of the quintic has three special
points: large volume point $z\to 0$ that we have already discussed in 
depth, the conifold singularity $z=5^{-5}$ at which the period $\calf_2$ 
vanishes, and the so-called Gepner or Landau-Ginzburg point, $z\to\infty$,
which is not a singularity of the CFT, but exhibits a $\zet_5$ orbifold
monodromy. We wish to understand the analytic behavior of $\calw$, or 
equivalently $\calt$, around each of these points. We shall work with the
ansatz
\begin{equation}
\eqlabel{ansatz}
\calt_\pm(z) = \frac{w^1(z)}2\pm\frac{w^2(z)}4\pm a \tau(z)
\end{equation}
and determine the coefficient $a$ from consistency requirements.

The standard tool to do the analytic continuation of solutions of a
hypergeometric differential equation of the type \eqref{opf} is the
Barnes integral representation. For $\tau$, this representation takes 
the form
\begin{equation}
\tau(z) = \frac{\pi^2}{60} \frac{1}{2\pi\ii}
\int_C \frac{\Gamma(-s+1/2)\Gamma(5s+1)\Gamma(s+1/2)}{\Gamma(s+1)^5}
\ee^{\ii\pi(s-1/2)} z^s
\end{equation}
where the integration contour is straight up the imaginary axis. For
$|z|<5^{-5}$, we close the contour on the positive real axis and recover
\eqref{tau}. For $|z|>5^{-5}$, we instead close the contour on the negative
real axis, and obtain the expansion
\begin{multline}
\eqlabel{small}
\tau(z) =\tau_1(z) + \tau_2(z) =
\frac{\pi^2}{60}
\Biggl[\sum_{m=0}^\infty \frac{-\Gamma(-5m-3/2)}{\Gamma(-m+1/2)^5} z^{-m-1/2}
 \\ + \sum_{m=1}^\infty 
\frac{-\Gamma(m/5) \ee^{4\pi\ii m/5}}{5\Gamma(m)\Gamma(1-m/5)^4} z^{-m/5} \; 
\ee^{-\ii\pi/2} \frac{\sin \pi m/5}{\cos \pi m/5}
\Biggr]
\end{multline}
The first term, $\tau_1(z)$, is simply the unique solution of \eqref{opf} with 
a squareroot behavior around $z=\infty$, and changes sign as we circle around 
$z^{1/5}\to\ee^{-2\pi\ii/5} z^{1/5}$. The second sum in \eqref{small} is easily 
verified to be a solution of the ordinary Picard-Fuchs equation, and hence a 
closed string period. To determine which one, we can compare it with the canonical 
$\zet_5$ symmetric basis of solutions of \eqref{picard} around the Gepner point 
\cite{cdgp}, ($j=0,\ldots,4$) 
\begin{equation}
\varpi_j(z) = \sum_{m=1}^\infty 
\frac{-\Gamma(m/5)\ee^{4\pi\ii m/5}}{5\Gamma(n)\Gamma(1-m/5)^4} z^{-m/5}
\; \ee^{2\pi\ii j m/5}
\end{equation}
Indeed, the identity
\begin{equation}
\frac{\sin\pi m/5}{\cos\pi m/5}
= 2 \sin 2\pi m/5-2\sin 4\pi m/5
\end{equation}
shows that
\begin{equation}
\tau_2(z) = \frac{\pi^2}{60}\bigl[\varpi_0+2\varpi_4+2\varpi_2\bigr]
\end{equation}
According to the results of \cite{cdgp}, the small volume period vector 
$\varpi = (\varpi_2,\varpi_1,\varpi_0,\varpi_4)^T$ is related to the large 
volume basis $\Ip=(\calf_1,\calf_2,w^1,w^2)^T$ via $\Ip = M\varpi$ with
\begin{equation}
M=\begin{pmatrix}
\frac{3}{5}& \frac{1}{5} & -\frac{21}{5}& -\frac{8}{5}\\
0& -1& 1& 0 \\
-\frac{1}{5}& -\frac{2}{5}& \frac{2}{5} & \frac{1}{5} \\ 
0& 0& -1& 0
\end{pmatrix} 
\end{equation}
This allows us to express $\tau_2(z)$ in the integral basis,
\begin{equation}
\eqlabel{integral}
\tau_2(z) = \frac{\pi^2}{60}\bigl[-4 \calf_1+8\calf_2-11w^1+15 w^2\bigr]
\end{equation}
Moreover, by using the known monodromy matrices around the Gepner point, 
we find that as $z^{-1/5} \to \ee^{2\pi \ii /5} z^{-1/5}$, 
\begin{equation}
w^1\to -\calf_2+w^1-w^2\,, \qquad
w^2\to \calf_2+w^2\,,\qquad
\tau\to -\tau +\frac{\pi^2}{60} \calf_2
\end{equation}
Thus we see that were it not for the quantum corrections of the domainwall
tension in \eqref{ansatz}, the Gepner monodromy would take $\frac{w^1}2+\frac{w^2}4$
to $\frac{w^1}2-\frac{w^2}4-\frac{\calf_2}4$, and would not induce a symmetry
of the domainwall spectrum as it should. Moreover, we see that the
lucky number that makes the Gepner monodromy integral is indeed 
$a=\frac{15}{\pi^2}$. (Strictly speaking, this is only the minimal 
possibility, a natural choice.) With this value, the Gepner monodromy acts as
\begin{equation}
\eqlabel{gepner}
A: \qquad \calt_+ \to \calt_-\,,\qquad\qquad
\calt_- \to \calt_+ - w^2 -\calf_2
\end{equation}
on the open string periods. Since as discussed in section \ref{supo},
the large volume monodromy acts by $T_\infty: \calt_-\to\calt_+$, 
$\calt_+\to\calt_-+w^2$, we find by combining the two that the conifold 
monodromy about $z=5^{-5}$, $T=T_\infty^{-1}\circ A^{-1}$ acts trivially 
on both $\calt_+$ and $\calt_-$. 

Let us verify this last assertion explicitly, in order to check that
everything is consistent. A straightforward way to compute this monodromy
is to compare the divergence of the large volume expansions 
\eqref{fundamental} and \eqref{tau} as $z$ approaches the singularity
$z\to z_*= 5^{-5}$. We know from \cite{cdgp} that at the conifold,
$\calf_2$ vanishes as $\calf_2\sim\alpha_1(z-z_*)+\alpha_2(z-z_*)^2+
\cdots$ and $\varpi_0$ behaves as $\varpi_0\sim \frac{1}{2\pi\ii} \calf_2
\log(z-z_*) + {\it regular}$. To determine the coefficient $b$ 
in $\tau\sim \frac{b}{2\pi\ii}\calf_2 \log(z-z_*) +{\it regular}$, we 
compare the second derivatives of $\varpi_0$ and $\tau$ as $z\to z_*$.
Using Stirling's formula, we find
\begin{equation}
\varpi_0'' \sim \sum_m (5^5z)^m \Bigl[\frac{5^{10}\sqrt{5}}{4\pi^2}
-\frac{7 \cdot 5^9 \sqrt{5}}{4\pi^2} \frac{1}{m} +\cdots \Bigr]
\end{equation}
which determines $\alpha_1$, $\alpha_2$. Doing the same for $\tau$
delivers
\begin{equation}
\tau'' \sim \sum_m (5^5z)^{m+1/2}\Bigl[\frac{5^{10}\sqrt{5}}{4\pi^2}
-\frac{7\cdot 5^9\sqrt{5}}{4\pi^2} \frac 1m +\cdots\Bigr]
\end{equation}
This implies $b=1$.

Thus, we find that the conifold monodromy takes $\tau\to\tau+
\frac{\pi^2}{60} \calf_2$, and since $w^2\to w^2-\calf_2$,
$\calt_\pm$ are invariant when we set $a=\frac{15}{\pi^2}$.

It is also worth pointing out that for $a=\frac{15}{\pi^2}$, the
leading behavior of $\calt_\pm$ as $z\to \infty$ is the same as 
that of an integral closed string period. This follows from \eqref{small} 
in conjunction with \eqref{integral}. As was mentioned in the
introduction, this is a further consistency check on our results.
It was shown in \cite{howa,strings} that the two open string vacua 
associated with the choice of discrete Wilson line (see subsection 
\ref{vacua}) could be identified with certain matrix factorization
in the Landau-Ginzburg B-model. At the Gepner point, $z\to\infty$,
the open string spectrum on the brane develops an extra massless 
state with a cubic superpotential. (This coalescence of open string 
vacua was first proposed in \cite{bdlr}.) There should therefore
be a domainwall between the two vacua that becomes tensionless as 
$z\to\infty$. Our result is then that while such a domainwall can 
indeed exist, it is not the most naive one obtained by wrapping a
D4-brane on the primitive disk, but has to be combined with the
appropriate integral period from \eqref{integral}.

To conclude this section, we summarize the results for the action of
the monodromies around Gepner point, conifold point, and large volume 
point on the extended period vector (we now use $\calt_-=-\calt_++w^1$)
\begin{equation}
\Ipp = \bigl( \calt_+, \calf_1, \calf_2, w^1, w^2 \bigr)^T
\end{equation}
We have:
\begin{equation}
\eqlabel{mondromies}
\begin{array}{ccc}
A & T & T_\infty\\\hline
\begin{pmatrix}
 -1& 0& 0& 1& 0\\
  0& 1& 3& 5& 3\\
  0& 1& -4& 8& -5\\
  0& 0& -1& 1& -1\\
  0& 0& 1& 0& 1
\end{pmatrix} 
&
\begin{pmatrix}
 1& 0& 0& 0& 0\\
 0& 1& 0& 0& 0\\
 0& 0& 1& 0& 0\\
 0& 0& 0& 1& 0\\
 0& 0& -1& 0& 1
\end{pmatrix}
&
\begin{pmatrix}
-1& 0& 0& 1& 1\\
 0& 1& 0& -5& -8\\
 0& -1& 1& -3& 5\\
 0& 0& 0& 1& 1\\
 0& 0& 0& 0& 1
\end{pmatrix}
\end{array}
\end{equation}
These matrices satisfy $A\cdot T\cdot T_\infty = 1$ and $A^{10}=1$,
but $A^5\neq 1$. Thus we find that the combined open-closed moduli 
space is a double cover of the quantum K\"ahler moduli space of the 
quintic, branched at $z=0$ and $z=\infty$.

\section{Localization in the A-model}
\label{localization}

In this section we shall show how to check the enumerative predictions 
that we have obtained using mirror symmetry. We have outlined the main 
strategy in the introduction, so we will attempt to be brief. Details
can be filled in from \cite{kontsevich} and \cite{horietal}, Chapter 27.

Consider the moduli space $\calm_d\equiv\overline{\calm}_{0,0}(\CP^4,d)$ 
of genus zero stable maps to $\CP^4$ in degree $d$. For each point 
$f:\Sigma\to\CP^4$ in $\calm$, we can pullback from $\CP^4$ the bundle 
$\calo(5)$ of quintic polynomials. The global sections of that bundle
$\calo(5d)$ over $\Sigma$ then fit together to a vector bundle $\cale_d$
as we vary $f$ over $\calm_d$. Any particular quintic polynomial 
$P(z_1,\ldots ,z_5)$ in the homogeneous coordinates of $\CP^4$ gives 
a section of $\calo(5)$. The resulting section of $\cale_d$ vanishes
at precisely those genus zero maps into $\CP^4$ which happen to be 
contained in the quintic given by $P$. This identifies the number of 
genus zero, degree $d$ maps to the quintic as the Euler class of $\cale_d$:
\begin{equation}
\eqlabel{euler}
\tilde N_d =\int_{\calm_d} c_{5d+1}(\cale_d)
\end{equation}
It was shown in \cite{kontsevich} that this Euler class can be very 
efficiently computed using Atiyah-Bott localization. The entire
structure described above carries an $(S^1)^5$ action inherited 
from the standard $U(5)$ action on $\CP^4$. On the homogeneous
coordinates, this torus acts as 
\begin{equation}
\eqlabel{action}
\TT^5 = (S^1)^5 \ni (\rho_1,\ldots,\rho_5) :[z_1:\cdots :z_5]
\mapsto [\rho_1z_1:\cdots:\rho_5 z_5]
\end{equation}
(This action can be complexified, of course, but we really only need the 
real torus.) On $\CP^4$, there are exactly five fixed points, $p_i$, of 
this torus action, defined by $z_j=0$, $j\neq i$. The fixed point loci on 
$\calm_d$ can be associated combinatorially with certain decorated tree
graphs, $\Gamma$. The vertices of these graphs (which can have arbitrary 
valence, ${\rm val}(v)$) correspond to (genus $0$) contracted components 
of the source $\Sigma$. They are labeled by one of the fixed points $p_i$ 
which tells where the component maps. The edges of the graph correspond
to non-contracted rational components of $\Sigma$ mapping onto the
coordinate line joining $p_i$ to $p_j$. They are labeled by a positive
integer $d$ describing the degree of that map. The constraints on this
decoration are that $p_v\neq p_{v'}$ for adjacent vertices $v$, $v'$
and that the sum of degrees on the edges be equal to the total degree
under consideration.

In general, the fixed loci are not isolated points, but consist of 
certain moduli spaces $\overline{\calm}_\Gamma$ arising from the contracted
components at the vertices (of valence $\ge 3$). One can then compute
the ($\TT^5$-equivariant) Euler class of the normal bundle of 
$\overline{\calm}_\Gamma$ inside of $\calm_d$, as well as the Euler 
class of $\cale_d$ at the fixed points. The integrals over the 
$\overline{\calm}_\Gamma$ can be done, and what results is a very 
explicit formula for $\tilde N_d$ given by a sum over graphs and 
labellings, divided by the appropriate symmetry factor.

We wish to accomplish something similar for the holomorphic maps of disks 
to the quintic with boundary on the real locus. 

As we have indicated before, any disk with boundary on the real locus 
can be completed to a sphere, and the two halves of that sphere 
contribute in the same relative homology class. Conversely, any
sphere of {\it odd} degree is cut in two by the real locus in a 
non-trivial one-cycle.\footnote{This is not true for even degrees: There 
can be real spheres of even degree without real points. In the real 
problem, they give rise to maps from the crosscap to the quintic. In other 
words, they will play a role in orientifolds. I am grateful to
Jake Solomon for extensive discussions on these issues.} Therefore, 
the number of disks of odd degree $d$ is equal to twice the number of 
spheres of degree $d$ which are invariant under complex conjugation of
source and target. On the real locus $\calm_d^\reals\subset\calm_d$,
complex conjugation defines a real structure on the bundle of quintics
$\cale_d$ (and, of course, on the tangent bundle). Since we are interested
in maps into a real quintic, we can identify the open Gromov-Witten 
invariant as \cite{jake}
\begin{equation}
\eqlabel{jake}
\tilde n_d = 2 \int_{\calm_d^\reals} \eul(\cale_d^\reals)
\end{equation}
In trying to apply localization to this problem, one is naively troubled 
by the fact that the torus action \eqref{action} does not commute with the
standard complex conjugation \eqref{conjugate}. However, it is easy to
realize that there is another real subtorus of $U(5)$ which does. This 
torus is two-dimensional and is the Cartan torus of $O(5)\subset U(5)$.
It is the natural four-dimensional analogue of the $S^1$ action used in
\cite{katzliu}. An equivalent way to describe this is to choose the 
alternative complex conjugation
\begin{equation}
\eqlabel{sigma}
\sigma:\quad[z_1:z_2:z_3:z_4:z_5] \mapsto 
[\bar z_2:\bar z_1:\bar z_4:\bar z_3:\bar z_5]
\end{equation}
which commutes with the subtorus $\TT^2$ of \eqref{action} defined by 
$\rho_2=\rho_1^{-1}$, $\rho_4=\rho_3^{-1}$, $\rho_5=1$. The nifty thing 
about this torus is that its fixed points on $\CP^4$ are identical to 
those of \eqref{action}. Moreover, it is not hard to see that the 
fixed points of $\TT^2$ acting on $\calm_d^\reals$ are simply those fixed 
points of $\TT^5$ acting on $\calm_d $ which are invariant under $\sigma$.

From this discussion, we see that our task is to take a real section of 
Kontsevich's calculation \cite{kontsevich} with respect to the complex 
conjugation $\sigma$. A moment's thought shows why this is feasible: Any 
$\sigma$-invariant decorated graph of odd total degree contains the real
locus of $\Sigma$ at the middle of an edge. In other words, the contracted
components of $\Sigma$ are away from the real locus. The upshot is that 
the integrals over the fixed loci are identical to those before.

To understand the Euler class of the normal bundle and of the bundle of 
real quintics, we are helped by the following elementary fact: If $V$ is 
any real vector bundle, then the square of its Euler class is the Euler 
class of its complexification,
\begin{equation}
\eqlabel{root}
\eul(V) = \sqrt{\eul(V\otimes\complex)}
\end{equation}
For bundles of high enough rank, this formula of course only makes sense
for the universal bundle, or in equivariant cohomology. The sign of the
squareroot in \eqref{root} is determined by the choice of orientation on
$V$ (which does not affect the canonical orientation of $V\otimes\complex$). 
In our situation, $\cale_d^\reals\otimes\complex=\cale_d|_{\calm_d^\reals}$ 
and since we already know $\eul(\cale_d)$, we are done.

Our graphical calculus is then very much as in \cite{kontsevich}. A 
moduli space of $\TT^2$-invariant disks corresponds to a tree graph
$\Gamma$ with vertices mapping to fixed points $p_{\mu(v)}$ 
(with $\mu(v)\in\{1,\ldots, 5\}$) and edges mapping to coordinate lines 
joining $p_i$ to $p_j$. There is one special vertex, call it the first 
one, on which ends an extra half-edge with odd degree, call it $d_0$. This 
restriction is to ensure that the total degree
\begin{equation}
d = d_0 + 2 \sum_{\rm edges} d(e) 
\end{equation}
can be odd. Another condition is that the special vertex cannot
map to $p_5$. This arises from the fact that when we reconstruct
a $\sigma$-invariant sphere by reflecting our graph on the half-edge,
the first vertex will be adjacent to its image, and $\sigma(p_5)=p_5$.

In taking a squareroot of the formulas in \cite{kontsevich}, we have to
fix the signs. In principle, this could be done by a careful analysis
such as advertised in \cite{jakethesis,jakecolumbia}. In practice, the 
condition that the answer be independent of the torus weights is enough 
to determine the sign. Explicitly, we have
\begin{multline}
\eqlabel{formula}
\int_{\overline{\calm}_\Gamma}\frac{\eul(\cale_d^\reals)}{\eul(N_\Gamma^\reals)}=
\prod_{\rm edges} 
\frac{\displaystyle\prod_{a=0}^{5d} \frac{a\lambda_i + (5d-a)\lambda_j}{d}}
{\displaystyle (-1)^d\frac{(d!)^2}{d^{2d}} (\lambda_i-\lambda_j)^{2d}
\prod_{\topa{k\neq i,j}{a=0}}^d\Bigl(\frac{a}d\lambda_i +
\frac{d-a}d\lambda_j-\lambda_k\Bigr)} \\\cdot
 \prod_{\rm vertices} \displaystyle\frac{1}{(5\lambda_v)^{{\rm val}(v)-1}}
\prod_{j\neq v}(\lambda_v-\lambda_j)^{{\rm val}(v)-1}
\cdot\biggl(\prod_{\rm flags}\frac{d}{\lambda_v-\lambda_j}\biggr)
\biggl(\sum_{\rm flags} \frac{d}{\lambda_v-\lambda_j}\biggr)^{{\rm val}(v)-3}
\\[.2cm] \cdot
\frac{\displaystyle\prod_{a=0}^{(5d_0-1)/2}
\frac{a\lambda_{\mu(1)}+(5d_0-a)\lambda_{\sigma(\mu(1))}}{d_0}}
{\displaystyle (-1)^{(d_0-1)/2}\frac{d_0!}{d_0^{d_0}}
(\lambda_{\mu(1)}-\lambda_{\sigma(\mu(1))})^{d_0}
\prod_{\topa{k\neq \mu(1),\sigma(\mu(1))}{a=0}}^{(d_0-1)/2}
\Bigl(\frac ad_0\lambda_{\mu(1)} + \frac{d_0-a}{d_0}\lambda_{\sigma(\mu(1))}
-\lambda_k\Bigr)}
\end{multline}
Here, it is understood that the torus weights satisfy $\lambda_2=
-\lambda_1$, $\lambda_4=-\lambda_3$, $\lambda_5=0$. Note that setting
$\lambda_5$ to zero introduces zero weight components in the above formula,
which however always exactly cancel between numerator and denominator.
In formula \eqref{formula}, it is also understood that in counting the 
valence of the vertex called $1$, the half edge counts full. 

The final formula is
\begin{equation}
\eqlabel{final}
\tilde n_d = 2 \sum_{\Gamma,\; {\rm labellings}} 
 \frac{1}{|\aut\Gamma|}\; \int_{\overline{\calm}_\Gamma}
\frac{\eul(\cale_d^\reals)}{\eul(N_\calm^\reals)}
\end{equation}
As in \cite{kontsevich}, $|\aut\Gamma|$ is the product of the order of the
automorphism group of $\Gamma$ as a decorated graph times the product of 
the degrees on the edges (including $d_0$). For the first few degrees, 
one reproduces the results from eq.\ \eqref{opinst} in section \ref{supo}.

\begin{acknowledgments}
My interest in this problem was revived when Jake Solomon told
me that the number of degree 1 holomorphic disks was $30$. I would
like to thank him for several helpful discussions and for sharing parts 
of his thesis. I am indebted to Katrin Wehrheim for patiently
explaining what could (and could not) be learned from FO$^3$. I would 
also like to thank Simeon Hellerman, Calin Lazaroiu, Wolfang Lerche, 
Andy Neitzke, Rahul Pandharipande, and Edward Witten for valuable 
discussions and Dan Freed and Frank Sottile for helpful correspondence. 
This work was supported in part by the DOE under grant number 
DE-FG02-90ER40542.
\end{acknowledgments}

\end{document}